\documentclass[default,iicol]{sn-jnl}

\jyear{2022}%

\theoremstyle{thmstyleone}%
\usepackage{mathtools}
\usepackage{caption}
\usepackage[utf8]{inputenc}
\usepackage{hyperref}
\usepackage[english]{babel}
\usepackage{amssymb}
\usepackage{amsfonts}
\usepackage{multirow}
\usepackage{tabularx}
\usepackage{xcolor}
\usepackage{amsmath}
\usepackage{booktabs}
\DeclareMathOperator*{\argmin}{\arg\min} 
\usepackage{rotating}
\usepackage{afterpage}

\usepackage{amssymb}
\usepackage{lipsum}

\newcommand{\rev}[1]{{#1}}

\newcommand{\revmax}[1]{{#1}}

\newcommand{\revfin}[1]{{#1}}

\theoremstyle{thmstyletwo}%

\theoremstyle{thmstylethree}%

\begin{document}
	
	\title[Article Title]{A Simulation--Based Optimization approach for analyzing the ambulance diversion phenomenon in an Emergency~Department network}
	
	\author[1]{\fnm{Christian} \sur{Piermarini}}\email{christian.piermarini@uniroma1.it}
	
	\author*[1]{\fnm{Massimo} \sur{Roma}}\email{roma@diag.uniroma1.it}
	
	\affil*[1]{\orgdiv{Dipartimento di Ingegneria Informatica, Automatica e Gestionale ``A. Ruberti''}, \orgname{SAPIENZA - Università di Roma}, \orgaddress{\street{via Ariosto 25}, \city{Roma}, \postcode{00185}, \country{Italy}}}

	\abstract{Ambulance Diversion (AD) is one of the possible strategies for relieving the worldwide  phenomenon of Emergency Department (ED) overcrowding. It can be carried out when an ED is overloaded and consists of redirecting incoming by ambulance patients to neighboring EDs. Properly implemented, AD should result in reducing delays of patient treatment, ensuring safety and rescue of life-threatening patients. From an operational point of view, AD corresponds to a resource pooling policy among EDs in a network. 
	\par\noindent
		In this paper we propose a novel model for studying the effectiveness of AD strategies, based on the Simulation--Based Optimization (SBO) approach. In particular,
		we developed a discrete event simulation model for reproducing the ED network operation.
		Then, for each AD policy considered, we formulate and solve an optimal resources allocation problem consisting of a bi-objective SBO problem where the target is the minimization of the non-value added time spent by patients 
		and the overall cost incurred by the ED network. A set of optimal points belonging to the Pareto frontier is obtained for each policy.
		To show the reliability of the proposed approach, a real case study consisting of six large EDs in the Lazio region of Italy is considered, analyzing the effects of \rev{adopting} different AD policies.}

	\keywords{Emergency Department overcrowding,
		Ambulance Diversion, Discrete Event Simulation, Simulation--based Optimization.}
	
	
	
	\maketitle

\newpage
\noindent
\textbf{Highlights}
\rev{
\begin{itemize}
	\item Our study focuses on Ambulance Diversion, a strategy that can be adopt by a network of EDs to reduce overcrowding.
	\item We propose a novel formulation that allows to determine an optimal resource allocation at each ED and to assess the adoption of different diversion policies. 
	\item The methodologies we use are based on discrete even simulation and black--box optimization.
	\item As result of our work, we provide ED managers and clinicians with a tool for coming up with the best resource distributions in an ED network, aiming at maximizing the efficiency of the service delivered and the utilization of medical staff and equipment.
\end{itemize}}
\newpage

\section{Introduction} \noindent
Among Emergency Medical Services (EMS), Emergency Department (ED) represents one of the most critical, since it provides first aid to patients arriving by ambulance or autonomously, sometimes for severe injures or even life--threatening. As well known, unfortunately, the global phenomenon of the overcrowding often negatively affects the services provided by an ED since it may cause delays in treatments, long waiting times and work overload of ED personnel. A wide literature has been devoted to deal with ED overcrowding and strategies to cope with such problem
(see e.g., \cite{Ahalt.2018,Bernstein.2003,Daldoul.2018,hoot.aronsky:08,Hoot.2007,Reeder.2003,Vanbrabant:20,Wang.2015,Weiss.2004,Weiss.2006}). 
\par
Here we focus on a strategy which \rev{can be} adopted aiming at obtaining an overall improvement of the services provided: the so called {\em Ambulance Diversion} (AD). It \revfin{can be adopted} when an ED is in an emergency situation and it consists of redirecting incoming by ambulance patients to neighboring EDs. Properly carried out, AD should result in relieving overcrowding and reducing delays of patient treatment, ensuring safety and rescue of life--threatening patients. From an operational point of view, AD corresponds to a resource pooling policy among EDs in a network of EMS. This implies a perspective change when studying patient flows, from a single ED to a network of EDs located in a particular geographical area.
\par
The first paper dealing with AD \cite{lagoe.1990} was published in \revfin{1990: AD} is proposed as a novel strategy for redirecting minor injured patients. In the 90's AD was a strategy rarely adopted \revfin{and} mainly in case of mass casualty accidents or critical events. In the 2000s, the adoption of AD policies became more frequent. This is clearly pointed out in the paper \cite{burt.2006} concerning the situation of EDs in the USA during 2003. In fact, the authors report that 
during the previous year, AD occurred in 45\% of the EDs, with an estimate of 501,000 ambulances diverted (one per minute). An insight study on AD in California \cite{abaris.2009},
confirmed the relevance of the AD phenomenon in this state, too.
\par
The growing interest of the Emergency Medicine international community for the AD is witnessed by several papers published in \revfin{the} literature (see, e.g., \cite{allon.2013,glushak.1997,hagtvedt2009cooperative,adkins.2015,gundlach.2010,lin.2015,kao.2015,baek2020centralized,tuller.2016,epstein.2006,pham.2006,ramirez.2009,ramirez.2011,ramirez.2011b,ramirez.2012,ramirez.2014,delgado.2013,li.2019,deo.2011,nakajima.2015}). In particular, we refer the reader to the recent review \cite{li.2019} (and the many references reported therein) which considers the more general problem of the ambulance offload delay, with a particular focus on AD (see Section~6.2.1 of \cite{li.2019}). In particular, in that paper the authors reviewed 137 papers and 58 of them deal with AD as solution of the ambulance offload delay. However, it can be easily observed that no more than 4 papers (among 58)  are published on Operations Research or Management Science journals. This clearly highlights the fact that a clinical point of view has been mainly adopted for analyzing the AD phenomenon, rather than focusing on more operational aspects. Of course, patient health and the related clinical aspects are the primary concerns to be considered, but an efficient management of AD can really play a crucial role in this regard.
\par
However, AD is the subject of a great deal of discussion, too. Indeed, several studies pointed out that not properly designed AD policies can even lead to a worsening of patient care (see, e.g., \cite{tuller.2016,delgado.2013,li.2019,deo.2011,nakajima.2015} and the references reported therein). The main drawback could be a delay in providing appropriate care to diverted patients, along with a number of others cons like ethical and legal ones.
As expected, studies evidenced the need of considering several factors besides a coarse index of the current ED overcrowding, mainly referring to patient severity conditions and status of neighboring EDs in a cooperative setting among EDs located in a geographical area. This underlines the guidelines issued by some national EMS systems to limit AD as much as possible or, at least, to reduce the length of time an ED \revfin{can be} on diversion status. On the other hand, it has been also pointed out that the limited use of AD may possible have negative effects on the overall EMS system (worsening ambulance offload delay and increasing of ED overcrowding) unless 
additional ED resources (human and/or structural) are provided (see, e.g., \cite{patel.2016,pforringer.2018} for analyses on \revfin{limitations} of the AD).
\par
The methodologies up to now used for studying AD phenomenon are both analytic and based on simulation models. In particular, in \cite{hagtvedt2009cooperative} the authors analyze cooperative strategies to reduce AD using many tools such as birth--death  \revfin{processes}, discrete event and agent--based simulation models, game theory. In \cite{ramirez.2014} AD optimal control policies are assessed by means of a Markov decision process, aiming at minimizing the average time that patients wait beyond the prefixed threshold value corresponding to their severity level. 
A queuing game between two EDs is introduced in \cite{deo.2011} to study the effects of decentralized and centralized AD settings; in the first case each ED has the objective of minimizing patient waiting times in the ED, while the average waiting time across all patients in the network is considered in the second case.
\par
However, the most commonly used methodologies for studying AD phenomenon are based on Discrete Event Simulation  (DES) models. In \cite{delgado.2013} a systematic literature review of studies on AD based on simulation models (updated to 2012) is reported. More recently, 
a DES model is proposed in \cite{lin.2015} aiming at evaluating different AD policies in terms of crowdedness index, patient waiting time before treatment and percentage of adverse patients, i.e. patients for which waiting time exceeds the threshold value admitted for their acuity level. A DES model is adopted in \cite{kao.2015} for studying the impact of AD on overcrowding in a regional ED network; results of an experimentation performed on EDs in a metropolitan region are also reported.
\par
A methodology that has been rarely adopted in studying the AD phenomenon, is Simulation--Based Optimization (SBO) (see, e.g., \cite{fu.2015,gosavi.2015}). This approach, whose
great potentialities are nowadays well known,
has been recently successfully adopted in different contexts of healthcare management (see, e.g., \cite{uriarte.2017,chen.2016,feng.2015,LMPRR:2016,LMPRR:2016b,Kuo.2023}). However, as far as the authors are aware, the only paper which reports the use of SBO methodology in dealing with the AD phenomenon is
\cite{ramirez.2011}. In this paper the authors combine a DES model with a multiobjective genetic optimization algorithm, aiming at selecting AD policies that minimize the expected time patients spend in non-value added activities, \revfin{namely} waiting and transportation. 
\par
\rev{In this paper, we propose a novel model for studying AD strategies, based on the SBO approach. In particular, we first develop a DES model reproducing the operation of a network of EDs, where each single ED is represented by a queueing system (input--throughput--output model).
	Then, we consider different AD policies and, for each of them, we formulate and solve an optimal resources allocation problem consisting of a bi-objective SBO problem with two conflicting objects: the minimization of the non-value added times (waiting times and possible transportation times due to AD) spent by patients 
	and the overall cost incurred by the ED network. A set of nondominated points belonging to the Pareto frontier is obtained for each policy. Such points give full information on objective function values and on objective trade-offs, providing the decision makers with the possibility to choose a specific Pareto optimal point most suited for their need. A comparison among different Pareto frontiers enables assessing the effectiveness of different AD policies and performing comparisons with respect to the case no AD strategy adopted.}
\par
\rev{To show the reliability of the proposed approach, we consider a real case study consisting of six large EDs in the Lazio region of Italy and we analyzed the effects of the adoption of different AD policies. The results obtained show that, under conditions of limited resources, if properly implemented, AD enables improving the performance of the ED network. Namely, the adoption of some AD policies \revfin{leads} to a significant reduction of patient waiting times before the first medical visit and treatment. Moreover, the experimentation performed also showed that not all AD policies are successful; indeed, some of them, based upon more restrictive conditions may be even detrimental with respect to the case no AD adopted.}
\par
\rev{The paper is organized as follows: in Section~\ref{sec:probstatement} we report a general framework of the features which characterize AD, highlighting the most common AD policies. Section~\ref{sec:singleEDmodel} describes the model we adopt for representing each single ED of the network. In Section~\ref{sec:statement} we report the SBO problem we state for dealing with AD in a network perspective and in Section~\ref{sec:implementation} we describe the implementation of the DES model we use for the network of EDs.
Section~\ref{sec:casestudy} deals with a real case study, where we apply the approach we propose to a network of six EDs. Finally, Section~\ref{sec:conclusions} includes some concluding remarks.}

\section{The Ambulance Diversion: terms and policies}\label{sec:probstatement}
Ambulance diversion is a practice adopted by EDs during overcrowding periods concerning patient arriving by ambulance. In fact, such a situation can lead to very long patient waiting times before treatment, hence increasing the risk of adverse events. 
In particular, if an ED can not deliver care for an incoming by ambulance patient due to resource unavailability, the following two strategies can be adopted:
\begin{itemize}
	\item patient is queued waiting for seat availability; in practice, paramedics continue to provide patient care in the ambulance or on a stretcher; 
	\item patient is transferred to another ED.
\end{itemize}
In the first case, possible ambulance offload delay also prevents the ambulance and its crew from returning to service. Therefore, besides long waiting time experienced by patient, a serious drawback is caused to the whole EMS, too. Therefore, to avoid such occurrence, ED managers can implement an AD strategy, namely incoming by ambulance patients can be {\em diverted} to a neighboring ED.
Of course, when implementing such a strategy, several factors must be taken into account, primarily patient acuity level and seat availability in other EDs.
Even if AD can not be considered a true solution to overcrowding, many National Healthcare Systems recommend adoption of AD to relieve the drawbacks due to crowded situations. 
\par\medskip
The main features which characterize AD are the following: \label{paginafeatures}
	\begin{itemize}
		\item the {\em Work Score} (WS) to be used for deciding if and when an ED has to go on AD status; 
		\item the {\em Initiating Criteria} (IC) adopted for selecting the best timing for an ED to initiates AD;
		\item the {\em Terminating Criteria} (TC) adopted for selecting the best timing for an ED to stopping AD;
		\item the {\em AD segment}, i.e. the time duration of the AD status before performing a new assessment of the WS is performed to possibly terminate AD (re-evaluation frequency);
		\item the {\em patient blocking rules} to be adopted for deciding which patients may be diverted;
		\item the {\em daily maximum number of hours allowed} for an ED to be on AD status;
		\item the {\em ambulance destination policies}, i.e. the criterion adopted for selecting among neighboring EDs which are more appropriate for receiving diverted patients. 
	\end{itemize}
The WS should be based on {\em ED crowding} and/or {\em staff workload}, enabling to keep track of ED conditions over time \cite{epstein.2006}. 
In most cases, the simple crowdedness index defined by the ratio between current loading of the ED and its full capacity is adopted, and exceeding a certain threshold value is used as AD initiating criterion \cite{lin.2015,kao.2015}. 
Other {\em single--threshold criteria} can be based on one of the following state variables often adopted as IC \cite{allon.2013,ramirez.2011b,ramirez.2010b}: 
\begin{enumerate}
	\item[-] number of available ED inpatient seats
	\item[-] number of patients waiting
	\item[-] number of patients boarding.
\end{enumerate}
``Patients waiting'' refers to patients that are in a queue waiting for the first medical visit. The term ``patient boarding'' refer to  patients holding in the ED after the decision that they must be admitted to an hospital ward has been taken; their stay in the ED (usually in an holding area) is due to unavailability of inpatients beds at the hospital. 
\par
Once an ED is on diversion, it keeps this status for a prefixed period (AD segment), at the end of which, a reassessment of the WS is performed: if conditions for AD are still satisfied, another AD segment is started, unless the daily maximum number of total hours allowed in a day for the AD is exceeded. Therefore, single--threshold criteria are usually characterized by three parameters \cite{ramirez.2011}: a threshold value on the state variable used for deciding when an ED initiates the AD status (IC); another threshold value on the same variable used for removing ED from the diversion status (TC); the re-evaluation frequency of the status for ED  currently on diversion, namely the AD segment; of course, diversion status can be removed only at a re--evaluation point. Note that periodic evaluation at discrete times is usually preferred with respect to a continuous review; indeed, this latter would require to monitor the state of the system \revfin{for each} event in the ED.
\par
As regards the blocking rules used by an ED which is on AD status, for selecting which patients can be possibly diverted, the most commonly considered are (see \cite{lin.2015,kao.2015}): 
\begin{itemize}
	\item[-] {\em all AD}, (A-AD), 
	i.e. all by ambulance arriving patients are diverted;
	\item[-]  {\em high-acuity AD}, 
	(H-AD), 
	i.e. only patients with severe injuries are diverted;
	\item[-]  {\em low-acuity AD},  (L-AD), 
	i.e. only patients with less severe injuries are diverted. 
\end{itemize}

\par
As regards the ambulance destination policies, the choice of the ED where to possibly transfer patients is usually based on the expected travel time between EDs and \revfin{on} the degree of overcrowding of the receiving ED. Therefore the destination policies usually adopted are: 
\begin{itemize}
	\item[-]  {\em the nearest ED},
	\item[-] {\em the least crowded ED}, regardless the distance between EDs..
\end{itemize}
Of course, patients can not be diverted towards EDs which are on AD status.

\section{{\em Input--throughput--output} simulation model for an ED}\label{sec:singleEDmodel}
\noindent 
The first step in studying a network of EDs is building of a model for a single ED. Since the focus is on the ambulance diversion phenomenon, it is appropriate to use, for each single ED, a model that summarizes the main care processes which contribute to overcrowding, without representing the detailed patient flow through the ED. In other words,
the aim is to observe the behaviour of an ED as a whole, without reproducing all the processes involved in ED operation. Hence we resort to a conceptual model commonly employed in this case, namely the so called \textit{input-throughput-output} simulation model. It 
is based on a {\em queueing model} where probability distributions which rule over patient flow through the ED are used and where parameters are calibrated by real data. Such model was introduced in \cite{Asplin.2003} for providing a practical framework for studying and possibly alleviate ED crowding. More recently, in \cite{lin.2015}, \cite{kao.2015} such model was adopted 
for studying the AD phenomenon. In these papers the authors show that by an appropriate calibration of the parameters, such a model well reflects the care processes that mostly contribute to ED overcrowding. 
\par
Based on this model, we represent the patient flow  through an ED as depicted in Figure~\ref{fig:conceptualmodel}:
\begin{figure*}[htbp]
	\centering
	\includegraphics[width=13.5cm]{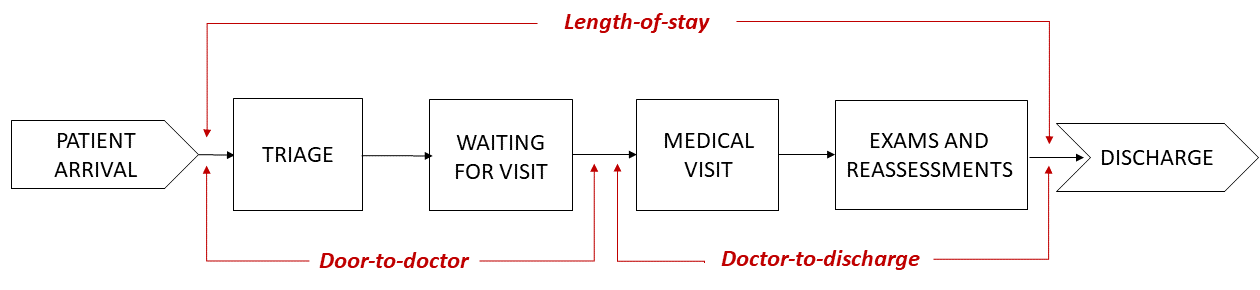}
	\caption{Conceptual model of a single ED of the network}\label{fig:conceptualmodel}
\end{figure*}
upon arrival, in the triage phase, a severity tag is assigned to each patient. Then patient wait for the first visit by a physician; after the visit, further examinations and reassessments can be carried out and patient is finally discharged from the ED. 
\par
Figure~\ref{fig:conceptualmodel} also reports two fundamental times: the time difference between patient arrival and the first visit, usually called \textit{door-to-doctor time} (DTDT) (see, e.g., \cite{vanbrabant2019simulation}), representing a critical waiting time especially for high--acuity patients; the time difference between the beginning of the first medical visit and the discharge from the ED named \textit{doctor-to-discharge time}. Of course, the sum of these two times provides the \textit{length-of-stay} (LOS).
\par
According to the standard assumption in the literature (see, e.g.,\cite{Whitt.2017,Kuo.2016,Zeinali.2015,Ahmed.2009,Ahalt.2018,Guo.2017,DGLMR:FSMJ}), we model patient arrival at the ED by a Nonhomogeneous Poisson Process (NHPP), approximating the arrival rate by a piecewise constant function. We consider that the service provided by the ED begins with the \revfin{first} medical visit and ends with the discharge; \revfin{we} assume that the corresponding service time, i.e. the doctor-to-discharge time, has a pattern specified by a suited probability distribution. However, unlike standard approaches (see, e.g., \cite{lin.2015}), we do not assume ``a priori'' a theoretical probability distribution for such time, but we obtain it by the real data of each single ED. Therefore, in each time interval where patient arrival rate is assumed constant, we model a single ED as an $M/G/s$ queueing system, i.e., a queue model with Poisson arrivals, service times having a general distribution and $s$ servers. It is worthy noticing that, when $s>1$, methods for computing the exact value of most performance metrics of this queueing model are not available.
\par
In this model a key role is played by the parameter $s$ which denotes the number of servers. In our model this parameter aims at capturing the total amount of resources available at an ED. It is an integer valued parameter symbolically representing both human and structural ED resources. Let us name this parameter \textit{sanitary resource}. A sanitary resource is seized by a patient when the visit begins and released at the discharge. In this manner, we do not need keeping track of resources of different type seized by a patient during the stay in the ED. A similar concept was introduced in \cite{lin.2015} for representing medical resources consumed by a patient.
\par
\begin{figure*}[htbp]
	\centering
	\includegraphics[width=14cm]{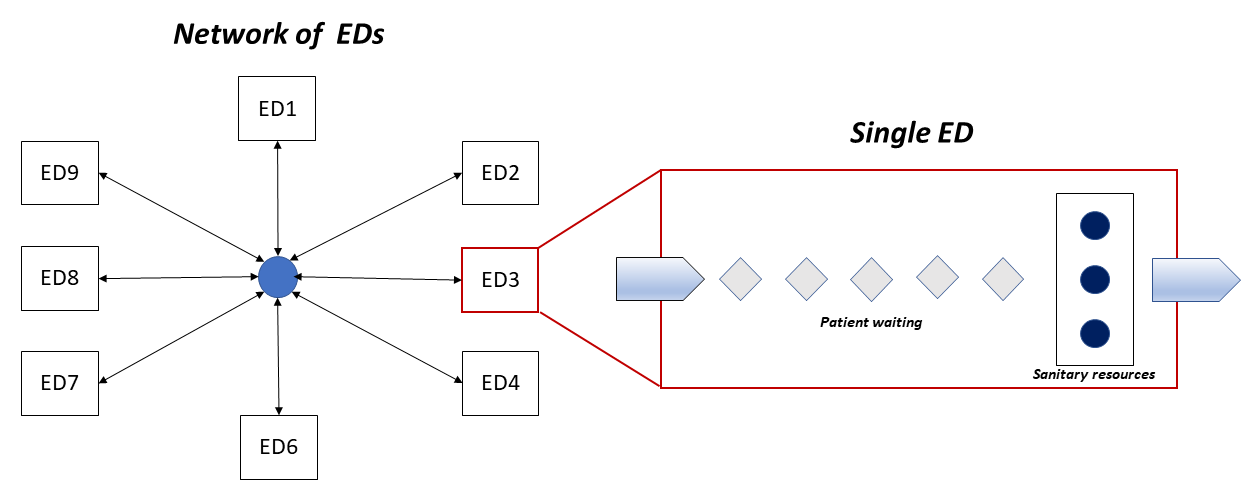}
	\caption{Example of a network of EDs and single ED queuing model}\label{fig:EDstructure}
\end{figure*}

The use of the sanitary resource concept allow us to overcome drawbacks usually present in other approaches when dealing with single ED model in a network. Examples of these disadvantages are: to relate the total amount of available resources at an ED only to its size or to compute these amount by means of empirical formulas; the need of defining a suited empirical scale for deciding the amount of resources required by a patient during the treatment (see, e.g., \cite{lin.2015,kao.2015}).
Figure~\ref{fig:EDstructure} depict a scheme of a network of EDs and the single ED queueing model that we adopt.
\par
We now describe in detail how we characterize the queueing system representing a single ED. For each ED \revfin{and for each severity tag} we assume that the following data are available:  
\begin{itemize}
	\item[-] the {\em patient interarrival times},
	\item[-] the {\em doctor-to-discharge times}.
\end{itemize}
Note that, even if many timestamps concerning the patient flow through an ED are usually unavailable from data (see, e.g., \cite{DGLMR:ANOR}), the timestamps that enable to compute these two times are usually registered.
\par
By using standard statistical procedure (see, e.g.,  \cite[Chapter~6]{law:15}), we first use these data to characterize the NHPP of patient arrival and to determine the probability distribution of the doctor-to-discharge times. 
\par
As concerns the first issue, in order to approximate the arrival rate by a piecewise constant function, the day hours are divided into time slots whose width can be fixed or varying (we refer the reader to \cite{DGLMR:FSMJ} for an insight discussion on how to determine the best piecewise constant approximation of the time-varying arrival
rate function, by finding the optimal partition of the day hours into a number of
not equally spaced intervals). For the sake of simplicity, here we consider fixed width \revfin{time} slots.
\par
As regards the second issue, based on summary statistics and histograms of the data, we hypothesize families of probability distributions that might be representative of the doctor-to-discharge times, determining the parameters via standard estimation procedures; then, we use goodness of fit tests for assessing whether the hypothesized probability distribution well represents real data. 
\par
To complete the definition of the input-throughput-output queueing model of a single ED, the value of the associated sanitary resource (i.e., the number of servers $s$) must be determined. In particular, to obtain a good accuracy in reproducing the operation of an ED, we divide the day hours into time slots and compute different probability distributions and the associated sanitary resource for each slot;
\revfin{it} is appropriate to match this time slots with staff shifts. In particular, once the probability distribution of the service time for a time slot has been defined, we adopt a {\em calibration procedure} for determining the most representative value of the sanitary resource. We now describe in detail such procedure.
\par
Calibration procedures are often used in dealing with ED DES models. This is due to the fact that, the input analysis needed for building an ED model is usually affected by missing data (see, e.g., the paper  \cite{vanbrabant:2019} for a discussion on quality of input data in ED simulation) thus requiring a proper estimation of some unknown parameters. The calibration procedure we propose aims at determining the value $s$ such that the KPIs of interest obtained from the simulation output are a good approximation of those derived from real data (see also \cite{DGLMR:ANOR}). In particular, we focus on the major KPI of interest when dealing with AD phenomenon, namely the door-to-doctor time. In fact, such KPI is one of the most critical to be considered since it strongly affects crowdedness indices and, as consequence, most of the WSs used for deciding if and when an ED must go on AD status. 
\par
To give a formalized description of the calibration procedure, let us denote by ${\cal T}$ the set of severity tags assigned at the triage.  Following the usual classification in five classes, like the Emergency Severity Index (ESI) adopted in the United States (see, e.g., \cite{Gilboy.2012}), it results ${\cal T}=\{1, \ldots , 5\}$, 
i.e., a scale from 1 (most urgent) to 5 (least urgent).
Moreover, we denote by $m$ the number of time slots we divide the day hours for defining  different probability distributions of the service time and the corresponding sanitary resource during the 24 hours. For each $j=1, \ldots , m$, we indicate by $s_j$ the sanitary resource associated to the {$j$-th} time slot. The calibration procedure consists of computing the sanitary resource $s_j$ as
\begin{equation}
	\argmin_{s_j} \sum_{k\in {\cal T}} \mid {\rm DTDT}^{j}_{k} - {\overline{\rm DTDT}}^j_k(s_j) \mid,
\end{equation}
where ${\rm DTDT}^j_k$ and ${\overline{\rm DTDT}}^j_k(s_j)$ denote the average door-to-doctor times for $k$-tagged patients in the $j$-th time slot obtained from data and from simulation output, respectively. 
In this manner, we estimate the values of the sanitary resources to be used in each single ED model (for each time slot), so that a good accuracy can be obtained in computing the average DTDTs from the simulation output. 
\par
Once the values of the sanitary resource has been determined for each ED in the network, a simulation model of the entire network can be constructed. It is represented by a weighted graph where nodes correspond to the EDs and edges to routes between EDs. This model reproduces the ``as--is'' status of the network, i.e., its current status without considering any AD: of course, all the analyses concerning the network of EDs \revfin{refer to this status as benchmark scenario}. In particular, following the ``what--if'' analysis approach, many scenarios could be reproduced aiming at evaluating how some KPIs vary based on the adoption of different AD strategies and different values of the sanitary resources. Instead, we are interested in a wider analysis which combines AD strategies and optimal allocation of the sanitary resources at each ED of the network. We give a detailed description of the approach we propose in the next section.


\section{Statement of a Simulation--based Optimization problem for AD in a network perspective}\label{sec:statement}
As we briefly recalled in the Introduction, AD must be analyzed in the right perspective: the  ``network perspective''. In fact, rerouting ambulances, to neighboring EDs should be a tool for balancing demand in an EMS network rather than only reducing overcrowding in single EDs. In other words, as well analyzed in \cite{deo.2011}, if the decision on diversion is taken by the managers of single EDs, their main aim is to alleviate crowding at their own ED; moreover their decision is based on observation of some local indicators, namely some WS referred to their own ED. As clearly evidenced by many studies (see, e.g., \cite{deo.2011} and the references reported therein), this could not have beneficial effects for patient health and ``centralized'' strategies should be adopted instead (see, e.g., \cite{baek2020centralized} \cite{ramirez.2011}).
\par
Based on this important observation, we propose a formulation of the AD problem based on the optimization of global KPIs, i.e., referred to the entire network. As pointed out in \cite{ramirez.2011}, the main goal of a centralized strategy for AD is the minimization of the time spent by patients in non-value added activities (NVA time),  namely, \textit{waiting time for the first medical visit (DTDT)} and \textit{transportation times due to diversion}. These times depend on the number of sanitary resources allocated at each ED in each time slot and on possible adoption of some AD policy. Of course, the greater the number of sanitary resources allocated in a time slot, the greater global management costs and the lower patient waiting times along with eventuality to go on diversion. 
Therefore, from a global perspective, we have two conflicting objects: from one hand, NVA times to be minimized and from the other hand, 
the costs due to sanitary resources to be minimized, too. Hence, we are faced with a multiobjective problem that must be solved by also considering the optimal allocation of resources at each ED of the network. 
\par
We focus on this multiobjective optimization problem which belongs to the class of simulation--based optimization problems, since objective functions and constraints are not available in analytic form, but obtained from runs of a DES model. Let us now formally define this problem.
\par
Assume that the network of EDs is composed of $n$ EDs located in an urban region. For each  $i=1, \ldots ,n$, assume that $m_i$ is the number of time slots adopted in the $i$-th ED when defining the probability distributions of the service times and the sanitary resources (see previous Section~\ref{sec:singleEDmodel}). For each $j=1, \ldots , m_i$, we denote by $s_{ij}$ the value of the sanitary resources allocated at the $i$-th ED in the $j$-th time slot, and by $d_{ij}$ the duration (in hours) of this time slot. Moreover, for each severity tag $k\in {\cal T}$ we denote by $T_{ij}^k$ the NVA times related to the $i$-th ED, in the $j$-th time slot, for $k$-tagged patients. Such times include patient waiting times at the $i$-th ED and any transportation times due to AD. In details, it results $T^k_{ij}=T^k_{ij}\left(s_{ij}; \xi(\omega)\right)$, i.e., it is function of $s_{ij}$ and \revfin{of} a random vector $\xi(\omega)$ representing the randomness. The random realizations $\xi_i$ of $\xi(\omega)$ correspond to different flow patient in the network.
\par
The bj-objective SBO problem we consider includes the following two objective functions to be minimized:
\begin{equation}\label{eq:f1}
	f_1(s_{ij})=\sum_{i=1}^n  \sum_{j=1}^{m_i} \sum_{k\in {\cal T}} \alpha_k \mathbb{E}\left[T_{ij}^k\left(s_{ij}; \xi(\omega)\right)\right],
\end{equation}

\begin{equation}\label{eq:f2}
	f_2(s_{ij})=\sum_{i=1}^n \sum_{j=1}^{m_i} d_{ij} s_{ij},
\end{equation}
where $\mathbb{E}[T_{ij}^k\left(s_{ij}; \xi(\omega)\right)]$ denotes the expected value of the times $T_{ij}^k\left(s_{ij}; \xi(\omega)\right)$ and
$\alpha_k\geq 0$ are suited weights.
\par
The first objective function consider the expected value of the NVA times for all the EDs of the network; the scalars $\alpha_k$  aims at, possibly, giving different weights to different severity tags. Of course, since it is computed by taking an expectation over a sample response function, it is not available in closed form. The objective function $f_2$ aims at quantifying the overall number of working hours provided by the allocated sanitary resources. The rationale behind the definition of this objective function is that the overall management costs are directly related to this quantity. 
\par
Moreover, we have constraints due to structural and regulatory restrictions. First, we have upper bound on the expected values $\mathbb{E}[T_{ij}^k\left(s_{ij}; \xi(\omega)\right)]$ due to threshold values $v_k$ established by National Healthcare Systems (NHS) for each severity tag, namely
\begin{equation}\label{eq:vinc1}
	\mathbb{E}[T_{ij}^k\left(s_{ij}; \xi(\omega)\right)] \leq v_k, \qquad k\in {\cal T},
\end{equation}
for all $i=1, \ldots , n$ and $j=1, \ldots , m_i$. Furthermore we have box constraints on the variables, namely
\begin{equation}
	l_{ij}\leq s_{ij} \leq u_{ij}, \quad i=1, \ldots ,n, ~ j=1, \ldots , m_i,
\end{equation}
where $l_{ij}$ and $ u_{ij}$ are integer non-negative lower and upper bound on $s_{ij}$ that are defined by each ED. 
\par
Hence, the bj-objective optimization problem we consider can be stated as 
\begin{equation}\label{eq:mosbo}
	\begin{aligned}
		& \min \bigl( f_1(s_{ij}), f_2(s_{ij}) \bigr)^\top &  \\
		& ~ {\rm s.t.} \quad \mathbb{E}[T_{ij}^k\left(s_{ij}; \xi(\omega)\right)] \leq v_k, \quad k\in {\cal T},\\ 
		& \qquad\qquad\qquad\qquad i=1, \ldots , n, \quad j=1, \ldots , m_i \\
		& \qquad\quad l_{ij}\leq s_{ij} \leq u_{ij},  \\ & \qquad\qquad\qquad\qquad i=1, \ldots , n, \quad
		  j=1, \ldots , m_i \\
		& \qquad\quad s_{ij} \quad \hbox{integer}.
	\end{aligned}
\end{equation}
Therefore, in this formulation the sanitary resources $s_{ij}$ are the decision variables of the optimization problem; the value of the sanitary resources determined by the calibration procedure described in the previous Section~\ref{sec:singleEDmodel} (corresponding to the ``as--is'' status) is used as starting point of the optimization algorithm adopted for solving problem \eqref{eq:mosbo}.
The aim is to find a trade-off between the reduction of the overall management cost, while guaranteeing patients timely treatments according to their severity tag.
\par
While the second objective function can be directly computed, the function $f_1$ can not be evaluated analytically, since it involves expected values of the NVA times; the same applies to constraints \eqref{eq:vinc1} and
this is a major issue when facing problem \eqref{eq:mosbo}. We adopt a commonly used technique for tackling this problem, namely the so called {\em sample average approximation} (see, e.g., \cite{Kleywegt.2001,KimSAA.2015}).
It consists of approximating the expected values of the sample response function with deterministic averages, namely
\begin{equation}\label{eq:saa}
	\mathbb{E}[T_{ij}^k\left(s_{ij}; \xi(\omega)\right)]\approx\frac{1}{N} \sum_{h=1}^N T_{ij}^k\left(s_{ij}; \xi_h\right),
\end{equation}
where $\xi_1 , \ldots , \xi_N$ is an independent identically distributed sample. Since we are using a DES model for reproducing the operation of the network of EDs, each sample corresponds to the output of a single replication of simulation runs and the average in \eqref{eq:saa} is computed over $N$ independent
replications. Therefore, the problem we are dealing with is a {\em Simulation--Based Multiobjective Integer Optimization problem}.
\par
Of course, given a sample $\xi_1, \ldots , \xi_N$, the right hand side of \eqref{eq:saa} is a deterministic function, hence, by using sample average approximation, (deterministic) black--box algorithms can by applied for solving problem \eqref{eq:mosbo}. However, usually standard optimization engines available within DES software packages can not deal with such a problem, hence an integration with an external implementation of a suited optimization algorithm may have to be necessary.
\par
It is important to note that, as well known (see, e.g., \cite{miettinen:99}), the solution of a bi-objective problem is provided in terms of \textit{Pareto frontier}, consisting of a set of \textit{Pareto optimal points} which represent a trade-off between the two conflicting objectives. They are nondominated points for which any feasible movement which improves function $f_1$
worsen the function $f_2$ and conversely. This fact represents an important merit of the approach we propose. Indeed, based on our approach we can provide the managers of the ED network with a set of optimal solutions (not only one optimal solution) and they can choose among them the most suited to their needs, still guarantying optimality.

\section{Implementation of the DES model of the network} \label{sec:implementation}
According to the description provided at the end of Section~\ref{sec:singleEDmodel}, we represented the network of EDs as a weighted graph where nodes correspond to the EDs and edges to routes between EDs. We implemented a DES model reproducing ambulance flow in the network and patient flow 
through each single ED. We used {\sf ARENA~16.1} simulation software by Rockwell Automation \cite{simulationwitharena}, a discrete event simulation package widely used also in healthcare management contexts. 
\par
A brief description of the implementation is the following. First, we constructed the single ED
input--throughput--output model described in Section~\ref{sec:singleEDmodel} (which is replicated at each node of the network). Model entities are patients (with an associated ambulance) and they are created as arrival entities at each ED, according to a suited NHPP. The {\sf\em severity$\_$tag$\_$attribute} is assigned to each entity on arrival according to the urgency. A check is performed on the variable {\sf\em ED$\_$status}: if it is {\sf\em on$\_$diversion}, a further check is carried out to verify if the adopted AD policy imposes patient to be diverted according to the selected \textit{patient blocking rules}; if so, patient is redirected to another ED following an \textit{ambulance destination policy}; otherwise, entity joins a priority queue waiting for the first medical visit. If {\sf\em ED$\_$status} is not {\sf\em on$\_$diversion}, entity directly joins the priority queue. When it is its turn and a sanitary resource is available, entity entries the {\sf\em visit\&treatments$\_$process} seizing one sanitary resource. Entity is delayed for a time depending on the probability distribution of the 
service time (doctor-to-discharge time) associated to the specific severity tag. Then entity releases the seized sanitary resource and exit the system.
\par
For each ED in the network an input analysis of available data is performed, as described in Section~\ref{sec:singleEDmodel}. In particular, from patient interarrival times, the NHPP of arrivals for each severity tag is determined and it is implemented by approximating the arrival rate by a piecewise constant function. From doctor-to-discharge times, for each time slot and for each severity tag, the probability distribution of the service times is defined and the value of the sanitary resources is computed by using the described calibration procedure.
To complete the DES model of the network, the routes between each ED and all the other ones are constructed and the corresponding transportation time is estimated from data concerning the road network.

\section{Experimentation on a case study}\label{sec:casestudy}
In this section we report the results of an experimentation performed on a real case study of a network of six EDs located in the urban area of Rome, Italy, which we named {\bf ED$_i$}, $i=1, \ldots, 6$.
The aim is to evaluate the reliability and the effectiveness of the approach we propose. 
\par
We collected data concerning EDs' operation for one year (before Covid-19 pandemic). The severity tag assigned at the triage was based on a four level (color) scale: red, yellow, green and white tags (from the most urgent to the least urgent). Our experimentation focus on high acuity cases, namely red and yellow tagged patients. The rationale behind this choice is twofold; first, for regulatory constraints,  usually AD may only concerns the most urgent cases, i.e., in this case, red and yellow tagged patients; secondly, high acuity patients use dedicated resources inside an ED, like specific treatment rooms  (Intensive Care Unit, Resuscitation Area) not shared with less urgent patients. 
\par
We performed the input analysis for both defining the NHPP of patient arrivals and determining the probability distributions of the service times of each ED of the network. To this aim, following the ED staff shifts, we adopt 8-hours time slots: \textit{(A)} from 00:00 to 08:00, \textit{(B)} from 08:00 to 16:00, \textit{(C)} from 16:00 to 24:00.
For the sake of simplicity, we adopt the same time slots for both approximating patient arrival rate with a piecewise constant function and for defining probability distributions of the service times.
\par
Table~\ref{tab:arrivals} in \revfin{the}  Appendix~\ref{sec:appendix} reports the number of yearly patient arrivals for each ED, for each time slot, for each red and yellow tagged patients. Note that red tagged patient arrivals in each time slot (and in particular during the night time slots) are few. Therefore, if we choose smaller time slots (e.g. 4-hour or 1-hour time slots) in estimating the patient hourly arrival rate, we would have even a smaller number of arrivals in each slot, hence a too small sample size to perform a significant statistical analysis. 
As regards the service times, by using standard statistics tools, we hypothesized probability distributions 
and estimated their parameters; then, by means of goodness of fit tests 
we measured how well the distributions fit the observations. 
Table~\ref{tab:input} in \revfin{the} Appendix~\ref{sec:appendix} reports the average hourly patient arrival rates (which define the piecewise constant function approximating the NHPP of patient arrivals) along with the probability distributions of the service times for each ED, for each time slot and for each red and yellow tag.
\par
The calibration procedure described in Section~\ref{sec:singleEDmodel} is then applied for determining the values of the sanitary resources which
reflect the current ``as--is'' status (without considering any AD) for each ED and for each time slot. The obtained values are reported in Table~\ref{tab:sanitaryresources} in the Appendix~\ref{sec:appendix}.
\par
To implement the network of these six EDs, we collected the average transportation times between EDs. Table~\ref{tab:matricetempi} in \revfin{the}  Appendix~\ref{sec:appendix} includes these times.
We are aware that a great variability of these times can be observed due to a number of factors. However, for this experimentation we used these average values since a more accurate estimate of these times would require an insight analysis of the traffic flow in the network not available at this stage of our study. 
\par
As concerns the design of experiments for our DES model, for each run we used 30 replications; the length of each replication is one year and the warm up period is 48 hours. \revmax{All the result statistics are computed on daily basis (24 hours) and times reported are expressed in minutes.}
\par
We then formulate the bi-objective SBO problem as stated in \eqref{eq:mosbo} with $n=6$ and $m_i=3$ for $i=1, \ldots , 6$. Moreover, following typical guidelines issued by Emergency Medicine authorities   concerning the ideal maximum time a patient in a severity category can wait for the first medical assessment, we set $v_{\tiny yellow}=15$ and  $v_{\tiny red}=5$ minutes (of course, in the case of red tagged patients, immediate assessment and treatment is recommended). In the definition of the objective functions we adopt $\alpha_k=1$ for all $k\in {\cal T}$ in \eqref{eq:f1} and $d_{ij}=8$, $i=1, \ldots ,6$, $j=1, \ldots ,3$ in \eqref{eq:f2}, corresponding to personnel shifts.
\par
From a simulation run we obtain the following values of the objective functions corresponding to the current ``as--is'' status reported in Table~\ref{tab:sanitaryresources}:
$f_1^0= 241.85$  and  $f_2^0=31,680$ minutes. It is important to note that, actually, this current status is infeasible for the bi-objective SBO problem \eqref{eq:mosbo}. Indeed, in some cases, constraints \eqref{eq:vinc1} are violated, i.e. the established threshold values are exceeded due to lack of resources. Therefore in our experiments we decided
to set the upper bounds $u_{ij}$ to a value greater than those reported in Table~\ref{tab:sanitaryresources}. Hence, 
we adopted the following values for the bounds on the sanitary resources:
$l_{ij}=2$ and $u_{ij}=6$, $i=1, \ldots , 6$, $j=1, \ldots ,3$. Of course, different values of these bounds could be considered for each ED and for each time slot.
\par
As concerns the efficient solution of the bi-objective SBO problem that we formulated for this experimentation, we leveraged on a Derivative-Free \revfin{Optimization} algorithm able to handle a multiobjective integer problem and which can provide us with good solutions, possibly requiring few iterations. In particular, we \revfin{used} the {\sf DFMOINT} method, recently proposed in \cite{DFMOINT} and freely downloadable from \textit{DFL--A Derivative--Free Library}\footnote{ URL: {\tt www.iasi.cnr.it/$\sim$liuzzi/DFL}}. In fact, it can face 
multiobjective optimization problems with both bound constraints on the variables and general nonlinear constraints, where objective and constraint
function values can only be obtained by mean of a black box. {\sf DFMOINT} is coded in Python and an interface with ARENA model is required. In particular, we also coded the SBO problem by using Python, creating an interface connecting ARENA with the external optimization engine. Starting from the value of the control variables $s_{ij}$ corresponding to the ``as--is'' status (see Table~\ref{tab:sanitaryresources}), at each iteration of the algorithm, the \revfin{value} of the control variables \revfin{is} updated and used as input \revfin{parameter} of the simulation. Then a simulation run is executed via the Python code by calling
executable ARENA files. The results are then passed to the optimization algorithm and a stopping criterion is checked. If it is verified, the procedure is terminated, otherwise another iteration is performed. We used the default stopping criterion of {\sf DFMOINT} algorithm, namely the maximum number iterations allowed that we set to 1,000. The obtained results provide a set of nondominated points, namely points belonging to the Pareto frontier (\textit{Pareto optimal points}).
\par
As a first experiment, we consider a starting scenario where AD is not allowed. This means that the NVA times $T_{ij}^k$ only include patient waiting times at the $i$-th ED since there are no transportation times.
That is, we solve the bi-objective SBO problem \eqref{eq:mosbo} without considering any AD. 
In Table~\ref{tab:sanitaryresourcesNO-AD} in \revfin{the} Appendix~\ref{sec:appendix}, we report the obtained Pareto optimal points along with the values of the functions $f_1$ and $f_2$ corresponding to these points. The first and the last row in this table correspond to the two extreme configurations: the first one refers to the lowest allocation of the sanitary resources and, consequently, the to the highest value of objective function $f_1$ (the NVA times); conversely, in the last one, the lowest value of $f_1$ is obtained due to the fact that the sanitary resources in all EDs except one are allocated at their maximum allowed value. All the intermediate configurations refer to other Pareto optimal points 
which represent a trade-off between the two extremes.
\begin{figure*}[htbp]
	\centering
	\includegraphics[width=12cm]{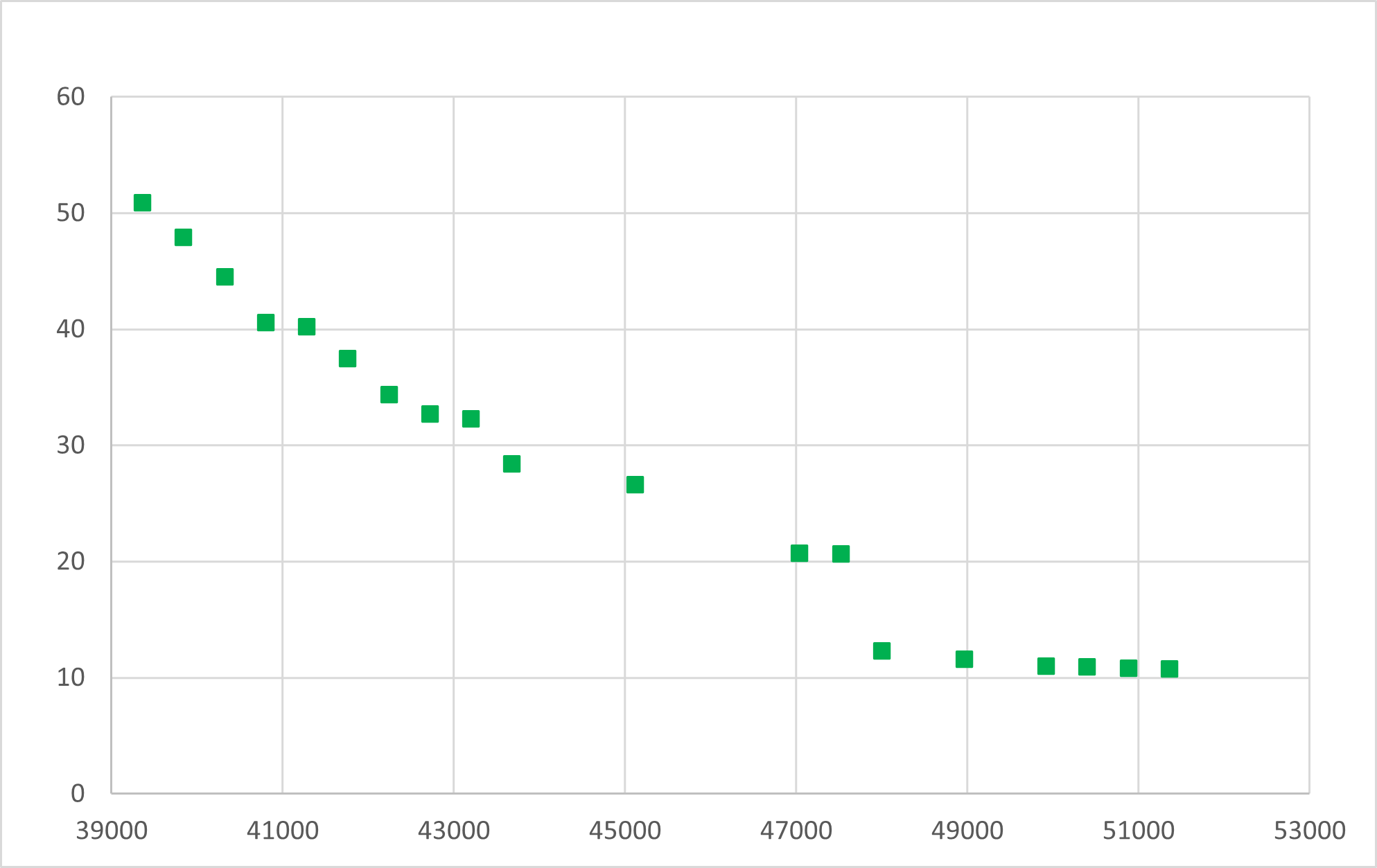}
	\caption{Pareto front obtained in the case no AD is considered. \rev{In the ordinate axis values of $f_1$ (in minutes); in the abscissa axis values of $f_2$ (in minutes)}}\label{fig:paretonodiv}
\end{figure*}
Figure~\ref{fig:paretonodiv} depicts in the objective space (i.e., the space
where the objective function values are used as coordinates) the obtained points of the Pareto frontier reported in Table~\ref{tab:sanitaryresourcesNO-AD}.
In the abscissa axis the values of \rev{the} objective function $f_2$ are reported, in the ordinate axis those of $f_1$. As it can be observed, the application of {\sf DFMOINT} algorithm enables obtaining a good coverage of the Pareto frontier. In the following we will use these results, which represent the optimal allocations of the resources in the case AD is not considered, as the benchmark for comparisons with scenarios involving AD.
\par
The scenarios concerning various AD policies we consider in our experimentation are based on different choices of the features characterizing AD reported at page~\pageref{paginafeatures}. Once a specific setting of features has been selected, we solve the associated bi-objective SBO optimization problem, determining the Pareto optimal solutions corresponding to the optimal resource allocation for this scenario.
For all the considered AD policies, we adopt as WS the number of sanitary resources and as IC the occurrence that all the sanitary resources are busy. As regards TC, AD status for an ED is stopped if, at the end of an AD segment, at least 1 sanitary resource has become available again.
As concerns \textit{patient blocking rule}, as previously described, we consider yellow and red tagged patients for possible diversion.
Therefore the diversion AD schemes we consider in our experimentation distinguish from each other in terms of \textit{AD segment length}, \textit{daily maximum number of hours allowed} for AD and  \textit{ambulance destination policies}. We tested several combinations of these features and, for the sake of brevity, we now reports results obtained for the following representative choices: 
\begin{itemize}
	\item \textit{AD segment length}: 6 hours long and 12 hours long;
	\item \textit{Daily maximum number of hours allowed} for the AD: 12 hour limit and no limit;
	\item \textit{Ambulance destination policies}: to the nearest ED and to the least crowded ED.
\end{itemize} 
Therefore we have 8 resulting different scenarios and, of course, some of them are more restrictive than others in the sense that they impose more tightening limitations. 
\par
\begin{figure*}[htbp]
	\centering
	\includegraphics[width=12cm]{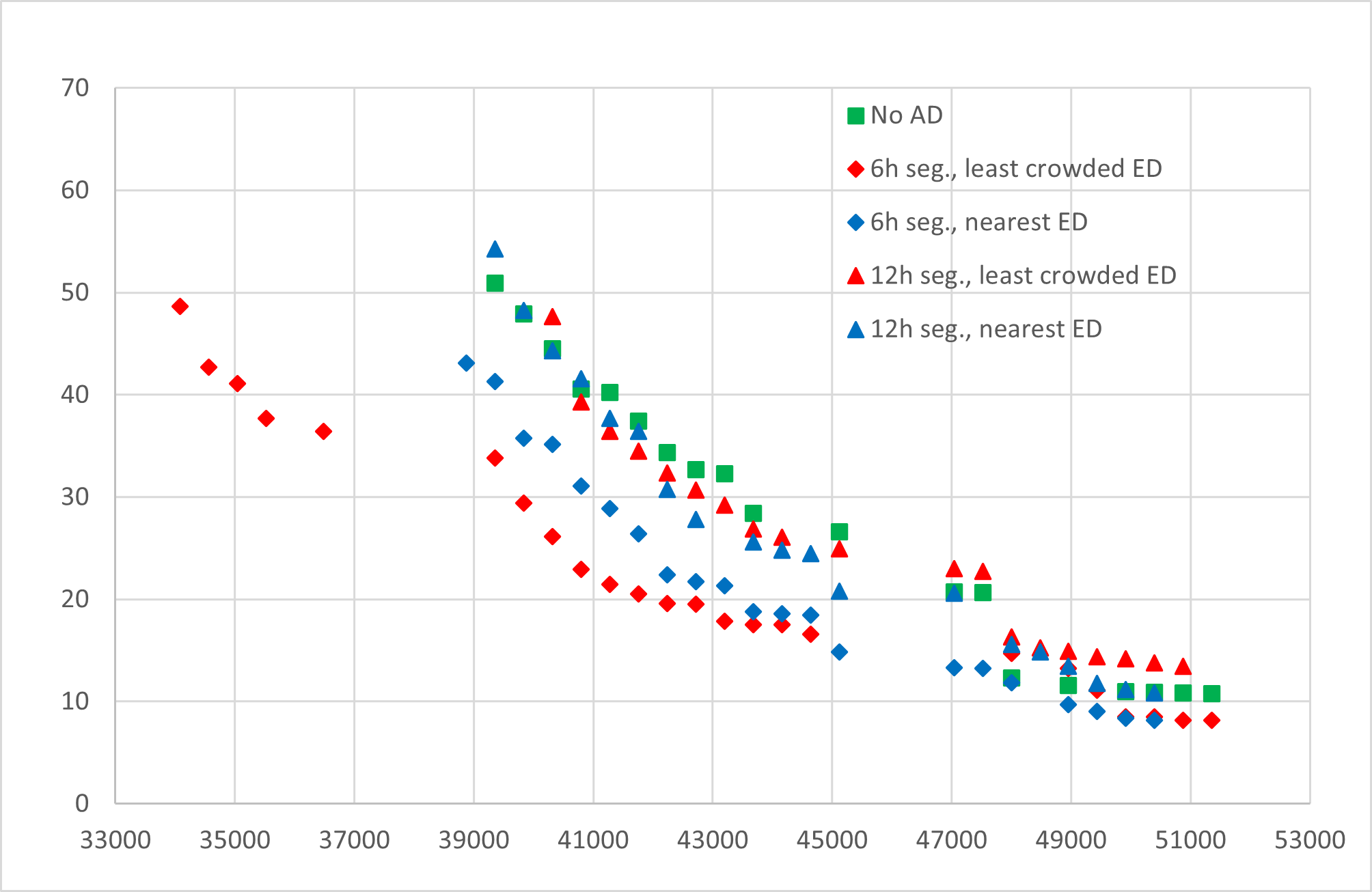}
	\caption{A subset of Pareto fronts referring to AD policies with a 12 hours limit on daily maximum number of AD hours allowed.}
	\label{fig:confronto1}
\end{figure*}
\begin{figure*}[htbp]
	\centering
	\includegraphics[width=12cm]{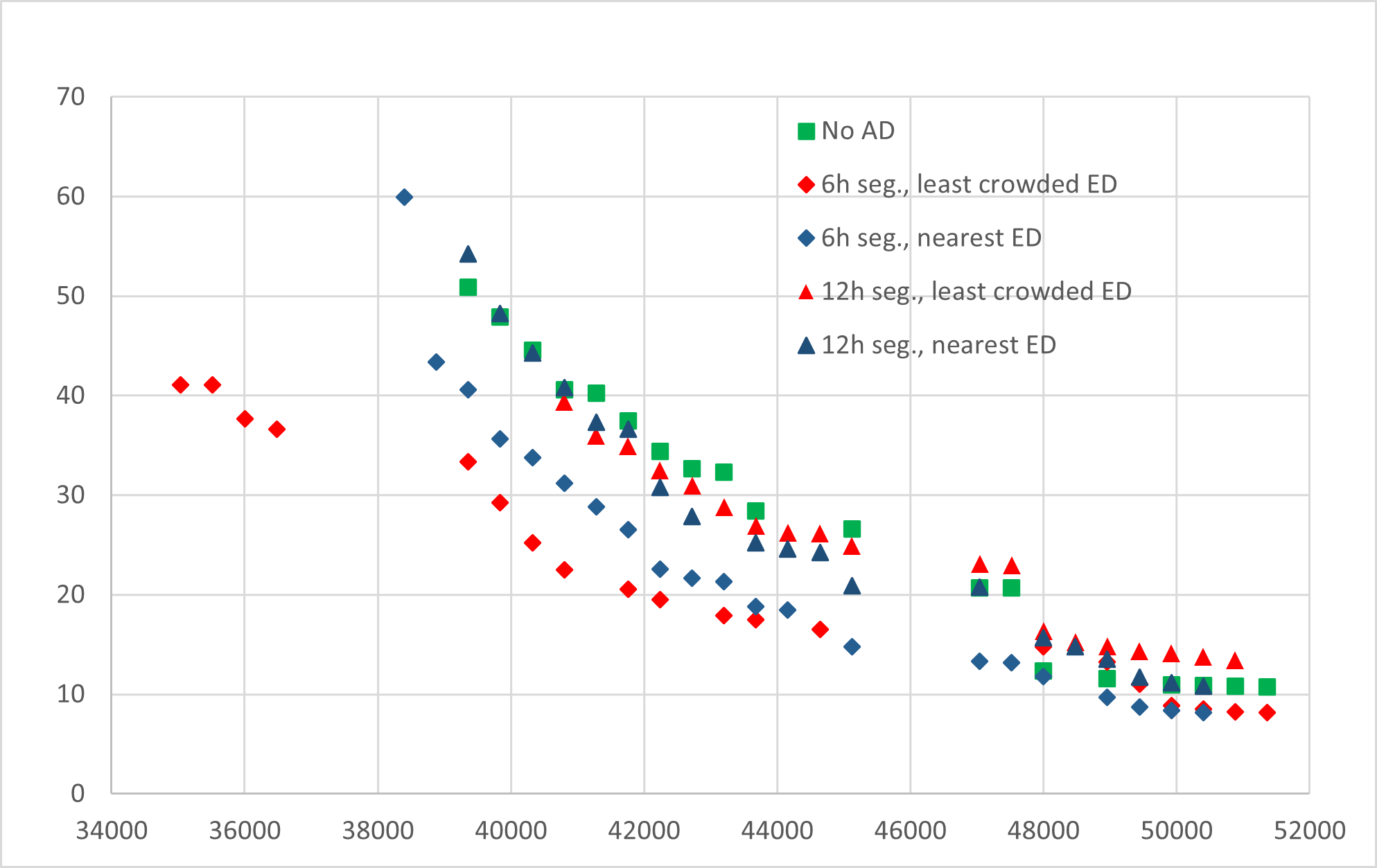}
	\caption{A subset of Pareto fronts referring to AD policies with no limit on daily maximum number of AD hours allowed.}
	\label{fig:confronto2}
\end{figure*}

Firstly, we compare the scenarios corresponding to a daily maximum number of 12 hours allowed with the case no AD is adopted. In particular, in Figure~\ref{fig:confronto1} the obtained
points of the Pareto frontier  are reported.
Then, in Figure~\ref{fig:confronto2} we report a similar comparison between no AD adopted and the scenarios in which no daily maximum number of hours limit is assumed.
Figure~\ref{fig:confronto1} and Figure~\ref{fig:confronto2} display some relevant outcomes obtained in our experiments. \rev{In fact, in both the figures, it can  easily be observed that the best policy is the one corresponding to the adoption of an AD segment of 6 hours and patients diverted towards the least crowded ED. Indeed, for values of $f_2$ (abscissa axis) less than 43,000 \revmax{minutes}, points belonging to the associated Pareto frontier (the red diamonds), dominates the points belonging to the other Pareto frontiers. Recalling that the objective function $f_2$ represents the overall \revmax{daily} \revmax{working time} provided by the allocated sanitary resources, this means that, when such number is below a certain value, the adoption of such AD policy leads to a significant improvement with respect to the no AD case. It can also be observed that, when this value falls below 38,000 \revmax{minutes}, then this policy is the only one providing Pareto optimal points. On the other hand, as expected, when the number of working \revmax{time} \revfin{provided by allocated sanitary resources} is high (say greater that 47,000 \revmax{minutes}) then the adoption of AD is even detrimental with respect to the case no AD is adopted.} 
\par
\rev{On the overall, the plots show that the worst AD policies are generally built upon the most restrictive conditions, i.e., 12 hours AD segments, 12 daily maximum diversion hours allowed and patient transferred towards the nearest ED. Observe that some of these policies
actually leads to no performance improvements or even to a worsening with respect to the no AD case. Adopting more flexible conditions enables obtaining significant improvements in terms of NVA times and number of necessary sanitary resources. This is confirmed by the fact that the best diversion policies \revfin{appear} to be the most flexible ones, imposing a small AD segment, no limit on the daily number of AD hours allowed and patient transportation to the least crowded ED, regardless the distance between EDs.} 

%
%
%
%
\par
\begin{figure*}[htbp]
	\centering
	\includegraphics[width=12cm]{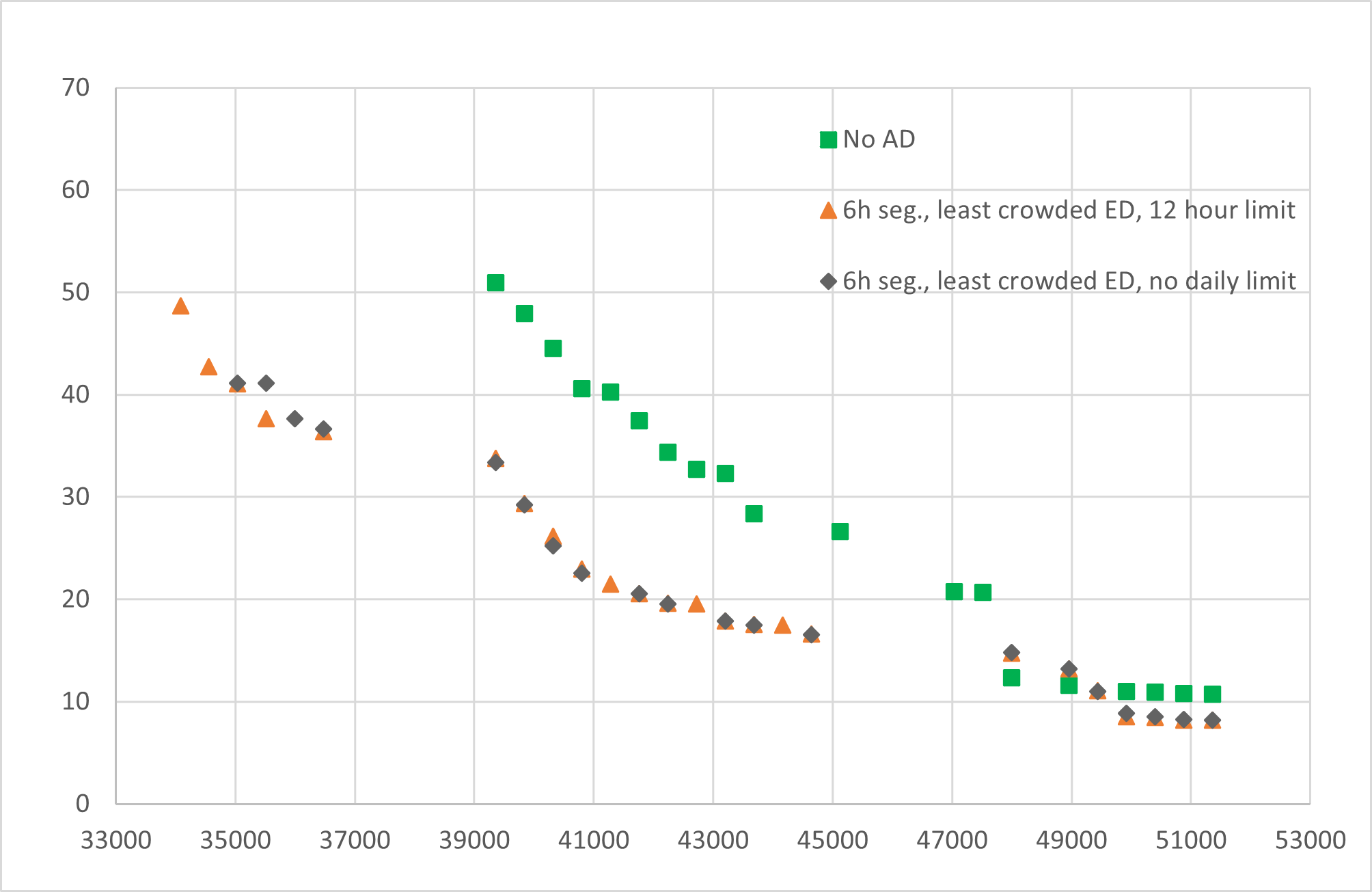}
	\caption{Comparison between the two best Pareto frontiers corresponding to: 6 hours segment, least crowded ED in the cases no limit on daily maximum number of hours allowed and 12 hours limit}
	\label{fig:confronto3}
\end{figure*}

\rev{A comparison between the optimal points of the Pareto frontier corresponding to the best policy obtained in Figure~\ref{fig:confronto1} and Figure~\ref{fig:confronto2} is reported in Figure~\ref{fig:confronto3}.}
\rev{The two Pareto frontiers corresponding to no daily time limit and to 12 hours daily time limit are almost overlapping. This indicates that imposing a limit on the daily AD hours allowed is not a crucial factor when pursuing optimal operative results of the network. Moreover, note that, since both curves refer to the use of 6 hours AD segment, the daily time limit of 12 hours coincides with a re-evaluation instant. Figure~\ref{fig:confronto3} also clearly evidences that, under conditions of limited resources (say $f_2$ less than 45,000 \revmax{minutes}), 
the points belonging to the Pareto frontier obtained when AD is not adopted are greatly dominated by the points of the Pareto frontier related to the two best AD policies. This show that, in this case, a great improvement can be obtained by adopting these AD policies.}
\par
\rev{Finally, in Table~\ref{tab:sanitaryresources-best-AD} in \revfin{the} Appendix~\ref{sec:appendix} we report the optimal values of the sanitary resources corresponding to the best AD policy of 6 hours AD segment, no limit on daily AD hours allowed and redirection towards the least crowded ED. By comparing this table with Table~\ref{tab:sanitaryresourcesNO-AD} (which refers to the case no AD adopted) it can be observed more in detail how this AD policy leads to a significant improvement of the performance of the ED network. Indeed, a direct comparison show as better (lower) values of the NVA times are obtained with a smaller number of sanitary resources allocated. Once again we highlight how providing ED managers with Pareto optimal points in place of a single optimal solution \revfin{enables} them to choose and adopt the strategy more appropriate to meet specific demands.}

\section{Conclusions}\label{sec:conclusions}
\noindent 
\rev{In this paper we propose a Simulation-Based Optimization approach for studying the AD phenomenon. The aim is to analyze the adoption of different AD strategies in a network of EDs, evaluating the effects on patient waiting time for the first visit by a physician. This approach performs an optimal resource allocation of each single ED and allows to assess possible advantages of patient interhospital transportations.}
\par
\rev{We propose a general formulation consisting of a Simulation--Based bi-objective Integer Optimization problem where the two objective functions to be minimized are: the expected value of patient waiting time for the first medical visit (also considering possible time for patient transportation due to diversion); the overall number of working hours provided by the allocated resources at each ED. 
We resort to a DES model for representing the operation of the network of EDs, where each ED consists of a suitably validated and calibrated input--throughput--output queueing model.}
\par
\rev{The solution of the bi-objective SBO problem is given in terms of a set of Pareto optimal points representing a trade-off between the required time to provide first medical care to urgent patients and the costs related to the allocated sanitary resources. This enables  
providing the decision makers with a set of optimal solutions to select according to their needs.}
\par
\rev{To assess the reliability and the effectiveness of our approach, we considered a first--aid network composed by six EDs located in the Lazio region of Italy. The real data collected for these EDs (related to one year) have been used for constructing a DES model reproducing the operation of the network. We implemented the DES model by using ARENA simulation software and we used a state-of-the-art DFO algorithm for solving the resulting bi-objective integer black--box optimization problem. The experiments we performed on this case study analyzed several diversion policies and their combinations. The results obtained highlighted that AD enables to greatly improving the performances of the ED network,
provided that the adopted policies are flexible enough. On the other hand, applying restrictive diversion policies could even lead to a worsening of the network performance with respect to no AD adopted; in such circumstances, AD should not be applied by any means.}
\par
\rev{In dealing with the real case study we adopted some simplifications imposed by lack of some data. The most relevant is represented by 
the fact that we do not consider the  great variability of patient transportation time between EDs. In fact, in our experimentation we used average times since a more accurate estimate of these times is not available at this stage of our study. Another limitation of our experimentation is the fact that we determine patient arrival process without distinguishing the weekdays and variability over successive weeks due to possible seasonal phenomena; this was necessary since \revfin{the available data refer only} to one year. Despite these limitations, we believe that our experimentation is reliable enough to allow us to draw trustworthy results.}
\par
\rev{On the overall, the approach we propose is general and it can be successfully applied for analyzing AD phenomenon in a network of EDs. The reliability of the obtained results certainly depends on the quality of the information available. Anyhow, the result of our work is a tool which can help both ED managers and physicians to come up with the best resource distributions in an ED network standpoint, aiming at maximizing the efficiency of the service delivered and the utilization of medical staff and equipment.}

\clearpage
\appendix
\section{Data and results of the experimentation on the case study}\label{sec:appendix}
In this Appendix, we include data and results of the experimentation concerning the case study reported in Section~\ref{sec:casestudy}.

\begin{table}[htbp]
	\caption{Number of yearly patient arrivals at each ED, for each time slot (\textit{A}, \textit{B}, \textit{C}), for each red and yellow tags}
	\label{tab:arrivals}
	\centering
	{\footnotesize
		\begin{tabular}{rl@{\hspace{.75cm}}lll@{\hspace{.75cm}}l}
			\toprule
			& \textit{Color tag} & {\em A} & {\em B} & {\em C} & \textit{Total} \\ \midrule
			\multirow{2}*{\textbf{ED$_1$}} &  \textit{yellow} & 875 & 2688 & 2279 & 5842 \\
			& \textit{red} & 94 & 199 & 187 & 480 \\ \midrule
			\multirow{2}*{\textbf{ED$_2$}} 	
			&  \textit{yellow} & 390 & 1592 & 1299 & 3281 \\
			& \textit{red}  & 56 & 153 & 107 & 316 \\ \midrule
			\multirow{2}*{\textbf{ED$_3$}} 	
			&  \textit{yellow} & 599 & 2331 & 1839 & 4769 \\
			& \textit{red} & 38 & 102 & 92 & 232 \\ \midrule
			\multirow{2}*{\textbf{ED$_4$}} 
			&  \textit{yellow} & 439 & 2118 & 1503 & 4060 \\
			& \textit{red} & 70 & 295 & 202 & 567 \\ \midrule
			\multirow{2}*{\textbf{ED$_5$}} 
			&  \textit{yellow} & 311 & 2381 & 1338 & 4003 \\
			& \textit{red} & 55 & 299 & 188 & 542 \\ \midrule
			\multirow{2}*{\textbf{ED$_6$}} 	
			&  \textit{yellow} & 324 & 1138 & 832 & 2294 \\
			& \textit{red} & 99 & 281 & 191 & 571 \\ 
			\bottomrule
	\end{tabular}}
\end{table}

\begin{table}[htbp]
	\caption{Values of the sanitary resources obtained by the calibration procedure for each ED and for each time slot (\textit{A}, \textit{B}, \textit{C}).}
	\label{tab:sanitaryresources}
	\centering
	{\footnotesize
		\begin{tabular}{c@{\hspace{.2cm}}c@{\hspace{.2cm}}cc@{\hspace{.2cm}}c@{\hspace{.2cm}}cc@{\hspace{.2cm}}c@{\hspace{.2cm}}cc@{\hspace{.2cm}}c@{\hspace{.2cm}}cc@{\hspace{.2cm}}c@{\hspace{.2cm}}cc@{\hspace{.2cm}}c@{\hspace{.2cm}}c}
			\toprule
			\multicolumn{3}{c}{\textbf{ED$_1$}} & \multicolumn{3}{c}{\textbf{ED$_2$}} & \multicolumn{3}{c}{\textbf{ED$_3$}} & \multicolumn{3}{c}{\textbf{ED$_4$}} & \multicolumn{3}{c}{\textbf{ED$_5$}} & \multicolumn{3}{c}{\textbf{ED$_6$}} \\
			\noalign{\smallskip}
			{\em A} & {\em B} & {\em C}& {\em A} & {\em B} & {\em C}& {\em A} & {\em B} & {\em C}& {\em A} & {\em B} & {\em C}& {\em A} & {\em B} & {\em C}& {\em A} & {\em B} & {\em C} \\
			\midrule
			4 & 4 & 4 & 4 & 5 & 4 & 4 & 4 & 3 & 4 & 5 & 2 & 4 & 5 & 2 & 3 & 3 & 2 \\
			\bottomrule			
	\end{tabular}}	
\end{table}

\par
\begin{table}[htbp]
	\caption{Average transportation times between EDs (in minutes)}
	\label{tab:matricetempi}
	\centering
	{\footnotesize
		\renewcommand\arraystretch{1.5}
		\begin{tabular}{c@{\hspace{.30cm}}|c@{\hspace{.15cm}}c@{\hspace{.15cm}}c@{\hspace{.15cm}}c@{\hspace{.15cm}}c@{\hspace{.15cm}}c}
			\toprule
			& \textbf{ED$_1$} &\textbf{ED$_2$}  & \textbf{ED$_3$} &\textbf{ED$_4$}  & \textbf{ED$_5$} & \textbf{ED$_6$} \\
			\midrule
			\textbf{ED$_1$}	& 0.00 & 42.59 & 29.32 & 28.40 & 19.32 & 30.58 \\
			\textbf{ED$_2$}    & 42.59 & 0.00 & 15.41 & 23.09 & 27.07 & 17.58\\
			\textbf{ED$_3$}	& 29.32 & 15.41 & 0.00 & 12.31 & 14.46 &  16.01\\
			\textbf{ED$_4$}	& 28.40 & 23.09 & 12.31 & 0.00 &7.04  & 9.45\\
			\textbf{ED$_5$}	&19.32  &27.07  &14.46  & 7.04 & 0.00 &17.18\\
			\textbf{ED$_6$}	& 30.58 & 17.58 & 16.01 & 9.45 & 17.18  &0.00\\
			\bottomrule
	\end{tabular}}
\end{table}

\begin{sidewaystable*} 
	\caption{Average hourly arrival rate and probability distributions of the service times (in minutes) for each ED, for each time slot (\textit{A}, \textit{B}, \textit{C}) and for each \textit{red} and \textit{yellow} tag.} 
	\label{tab:input}
	\centering
	{\footnotesize
		\begin{tabular}{l@{\hspace{.2cm}}rllll}
			\toprule
			&  &  \multicolumn{2}{l}{ \textbf{Arrival rate}}& \multicolumn{2}{l}{\textbf{Service time}}  \\ 
			&  & \textit{red tag} & \textit{yellow tag} & \textit{red tag} & \textit{yellow tag}   \\ 
			\midrule
			\multirow{3}*{\textbf{ED$_1$}} &  \textit{A} & 0.068 & 0.920 & 0.999 + 5.86e+03 $\cdot$ BETA(0.113, 2.24)&0.999 + EXPO(197) \\
			& \textit{B} &0.064 &0.780&0.999 + WEIB(135, 0.586)&0.999 + 2.69 $\cdot$ BETA(1.25, 18.7) \\
			& \textit{C} &0.032 &0.303& 0.999 + 935 $\cdot$ BETA(0.361, 1.12)&0.999 + GAMM(198, 0.913)   \\ 	\hline\noalign{\smallskip}	
			\multirow{3}*{\textbf{ED$_2$}} & \textit{A} & 0.052&0.545 &0.999 + EXPO(253)& 0.001 + 4.3e+03 $\cdot$ BETA(1.58, 25.2)\\
			& \textit{B} &0.036 &0.444&0.999 + WEIB(234, 0.769) &0.999 + LOGN(300, 1.01e+03) \\
			& \textit{C} &0.019 &0.133& 0.001 + WEIB(238, 0.743)&0.999 + 2.35e+03 $\cdot$ BETA(0.594, 5.66) \\ 	\hline\noalign{\smallskip}	
			\multirow{3}*{\textbf{ED$_3$}} & \textit{A} &0.034 &0.798&0.001 + EXPO(185) & 0.001 + 2.52e+03 $\cdot$ BETA(1.28, 10.5)\\
			& \textit{B} &0.031 &0.629&0.999 + EXPO(176) & 0.001 + EXPO(266)\\
			& \textit{C} &0.013 &0.205 &1 + LOGN(333, 1.18e+03)& 0.001 + LOGN(384, 706)\\ 	\hline\noalign{\smallskip}	
			\multirow{3}*{\textbf{ED$_4$}} &  \textit{A} & 0.101&0.725 &0.999 + EXPO(163)&0.999 + 2.87e+03 $\cdot$ BETA(1.12, 13.7) \\
			& \textit{B} &0.069 &0.515&0.999 + WEIB(112, 0.67) &0.999 + EXPO(184) \\
			& \textit{C} &0.023 &0.150 &0.999 + EXPO(181)&2.12e+03 $\cdot$ BETA(0.638, 5.33) \\ 	\hline\noalign{\smallskip}	
			\multirow{3}*{\textbf{ED$_5$}} & \textit{A} &0.096 &0,815 &0.001 + EXPO(159)& 0.999 + 2.87e+03 $\cdot$ BETA(1.12, 13.7) \\
			& \textit{B} &0.065 &0,458 &0.001 + WEIB(109, 0.662)& 0.999 + EXPO(184)\\
			& \textit{C} &0.034 & 0,106&0.001 + EXPO(174)&0.999 + 2.12e+03 $\cdot$ BETA(0.632, 5.3) \\ 	\hline\noalign{\smallskip}	
			\multirow{3}*{\textbf{ED$_6$}} & \textit{A} &0.019 & 0.389&8 + WEIB(193, 0.567)&0.999 + 1.58e+03 $\cdot$ BETA(1.32, 10.4) \\
			& \textit{B} &0.024 &0.285 &WEIB(205, 0.935) &0.999 + GAMM(134, 1.17) \\
			& \textit{C} &0.012 &0.111 &17 + EXPO(210)&0.999 + WEIB(213, 1.16) \\
			\bottomrule
	\end{tabular}}	
\end{sidewaystable*}

\begin{table*}[htbp]
	\caption{Optimal values of the sanitary resources obtained for each ED and for each time slot (\textit{A}, \textit{B}, \textit{C}), in the case AD is not adopted.}
	\label{tab:sanitaryresourcesNO-AD}
	\centering
	{\footnotesize
		\begin{tabular}{c@{\hspace{.2cm}}c@{\hspace{.2cm}}cc@{\hspace{.2cm}}c@{\hspace{.2cm}}cc@{\hspace{.2cm}}c@{\hspace{.2cm}}cc@{\hspace{.2cm}}c@{\hspace{.2cm}}cc@{\hspace{.2cm}}c@{\hspace{.2cm}}cc@{\hspace{.2cm}}c@{\hspace{.2cm}}ccc}
			\toprule
			\multicolumn{3}{c}{\textbf{ED$_1$}} & \multicolumn{3}{c}{\textbf{ED$_2$}} & \multicolumn{3}{c}{\textbf{ED$_3$}} & \multicolumn{3}{c}{\textbf{ED$_4$}} & \multicolumn{3}{c}{\textbf{ED$_5$}} & \multicolumn{3}{c}{\textbf{ED$_6$}}
			& \multirow{2}{*}{\normalsize{$f_1^*$}} & \multirow{2}{*}{\normalsize{$f_2^*$}}\\
			\noalign{\smallskip}
			{\em A} & {\em B} & {\em C}& {\em A} & {\em B} & {\em C}& {\em A} & {\em B} & {\em C}& {\em A} & {\em B} & {\em C}& {\em A} & {\em B} & {\em C}& {\em A} & {\em B} & {\em C} &&\\
			\midrule
			5	&	5	&	3	&	5	&	5	&	5	&	5	&	5	&	3	&	5	&	5	&	5	&	5	&	5	&	5	&	3	&	5	&	3 &	50.92  &	39,360 \\
			5	&	5	&	3	&	5	&	5	&	5	&	5	&	5	&	3	&	5	&	5	&	5	&	5	&	5	&	4	&	5	&	5	&	3 &	47.90  &	39,840 \\
			5	&	5	&	3	&	5	&	5	&	5	&	5	&	5	&	4	&	5	&	5	&	5	&	5	&	5	&	4	&	5	&	5	&	3 &	44.53  &	40,320 \\
			5	&	5	&	4	&	5	&	5	&	5	&	4	&	5	&	5	&	5	&	5	&	5	&	5	&	5	&	4	&	5	&	5	&	3 &	40.57  &	40,800 \\
			5	&	5	&	4	&	5	&	6	&	5	&	5	&	5	&	3	&	5	&	5	&	5	&	5	&	5	&	5	&	5	&	5	&	3 &	40.24  &	41,280 \\
			5	&	5	&	4	&	5	&	6	&	5	&	5	&	5	&	4	&	5	&	5	&	5	&	5	&	5	&	5	&	5	&	5	&	3 &	37.45  &	41,760 \\
			5	&	5	&	5	&	6	&	6	&	5	&	3	&	5	&	5	&	5	&	5	&	5	&	5	&	5	&	5	&	5	&	5	&	3 &	34.38    &	42,240 \\
			5	&	5	&	5	&	6	&	6	&	5	&	5	&	5	&	5	&	5	&	5	&	5	&	5	&	5	&	4	&	5	&	5	&	3 &	32.69  &	42,720 \\
			5	&	5	&	5	&	6	&	6	&	5	&	5	&	5	&	4	&	5	&	5	&	5	&	5	&	5	&	5	&	5	&	5	&	4 &	32.31  &	43,200 \\
			5	&	5	&	5	&	6	&	6	&	5	&	5	&	5	&	5	&	5	&	5	&	5	&	5	&	6	&	5	&	5	&	5	&	3 &	28.39  &	43,680 \\
			6	&	6	&	5	&	6	&	6	&	5	&	5	&	5	&	5	&	5	&	5	&	5	&	5	&	5	&	5	&	5	&	5	&	5 &	26.62  &	45,120 \\
			6	&	6	&	4	&	6	&	6	&	6	&	4	&	6	&	6	&	6	&	6	&	6	&	6	&	6	&	4	&	6	&	6	&	2 &	20.72  &	47,040 \\
			6	&	6	&	6	&	6	&	6	&	6	&	4	&	5	&	6	&	6	&	6	&	6	&	6	&	6	&	4	&	6	&	6	&	2 &	20.66  &	47,520 \\
			6	&	6	&	4	&	6	&	6	&	6	&	4	&	6	&	6	&	6	&	6	&	6	&	6	&	6	&	6	&	4	&	6	&	4 &	12.31  &	48,000 \\
			6	&	6	&	4	&	6	&	6	&	6	&	4	&	6	&	6	&	6	&	6	&	6	&	6	&	6	&	6	&	6	&	6	&	4 &	11.59      &	48,960 \\
			6	&	6	&	6	&	6	&	6	&	5	&	5	&	6	&	6	&	6	&	6	&	5	&	6	&	6	&	6	&	6	&	6	&	5 &	10.97  &	49,920 \\
			6	&	6	&	6	&	6	&	6	&	6	&	5	&	6	&	6	&	6	&	6	&	5	&	6	&	6	&	6	&	6	&	6	&	5 &	10.91  &	50,400 \\
			6	&	6	&	6	&	6	&	6	&	6	&	5	&	6	&	6	&	6	&	6	&	5	&	6	&	6	&	6	&	6	&	6	&	6 &	10.81   &	50,880 \\
			6	&	6	&	6	&	6	&	6	&	6	&	5	&	6	&	6	&	6	&	6	&	6	&	6	&	6	&	6	&	6	&	6	&	6 &	10.75  &	51,360 \\
			\bottomrule			
	\end{tabular}}	
\end{table*}
\renewcommand\arraystretch{1}

\begin{table*}[htbp]
	\caption{Optimal values of the sanitary resources obtained for each ED and for each time slot (\textit{A}, \textit{B}, \textit{C}), in the case the best AD policy (6 hours AD segment, no limit on daily AD hours and redirection towards the least crowded ED) is adopted.}
	\label{tab:sanitaryresources-best-AD}
	\centering
	{\footnotesize
		\begin{tabular}{c@{\hspace{.2cm}}c@{\hspace{.2cm}}cc@{\hspace{.2cm}}c@{\hspace{.2cm}}cc@{\hspace{.2cm}}c@{\hspace{.2cm}}cc@{\hspace{.2cm}}c@{\hspace{.2cm}}cc@{\hspace{.2cm}}c@{\hspace{.2cm}}cc@{\hspace{.2cm}}c@{\hspace{.2cm}}ccc}
			\toprule
			\multicolumn{3}{c}{\textbf{ED$_1$}} & \multicolumn{3}{c}{\textbf{ED$_2$}} & \multicolumn{3}{c}{\textbf{ED$_3$}} & \multicolumn{3}{c}{\textbf{ED$_4$}} & \multicolumn{3}{c}{\textbf{ED$_5$}} & \multicolumn{3}{c}{\textbf{ED$_6$}}
			& \multirow{2}{*}{\normalsize{$f_1^*$}} & \multirow{2}{*}{\normalsize{$f_2^*$}}\\
			\noalign{\smallskip}
			{\em A} & {\em B} & {\em C}& {\em A} & {\em B} & {\em C}& {\em A} & {\em B} & {\em C}& {\em A} & {\em B} & {\em C}& {\em A} & {\em B} & {\em C}& {\em A} & {\em B} & {\em C} &&\\
			\midrule
			4 & 5 & 5 & 4 & 4 & 4 & 4 & 4 & 4 & 4 & 4 & 4 & 4 & 4 & 4 & 4 & 4 & 3 & 41.11 & 35,040 \\ 
			4 & 5 & 5 & 4 & 4 & 4 & 4 & 4 & 4 & 4 & 4 & 4 & 4 & 4 & 4 & 4 & 4 & 4 & 41.09 & 35,520 \\ 
			5 & 5 & 5 & 4 & 4 & 4 & 4 & 4 & 4 & 4 & 4 & 4 & 4 & 4 & 4 & 4 & 4 & 4 & 37.65 & 36,000 \\ 
			5 & 5 & 6 & 4 & 4 & 4 & 4 & 4 & 4 & 4 & 4 & 4 & 4 & 4 & 4 & 4 & 4 & 4 & 36.61 & 36,480 \\ 
			5 & 5 & 5 & 5 & 5 & 5 & 5 & 5 & 5 & 5 & 5 & 5 & 5 & 5 & 5 & 2 & 3 & 2 & 33.34 & 39,360 \\ 
			5 & 5 & 5 & 5 & 5 & 5 & 5 & 5 & 3 & 5 & 5 & 5 & 5 & 5 & 3 & 5 & 5 & 2 & 29.23 & 39,840 \\ 
			5 & 5 & 5 & 5 & 5 & 5 & 5 & 5 & 3 & 5 & 5 & 3 & 5 & 5 & 5 & 5 & 5 & 3 & 25.22 & 40,320 \\ 
			5 & 5 & 5 & 5 & 5 & 5 & 5 & 5 & 3 & 5 & 5 & 5 & 5 & 5 & 4 & 5 & 5 & 3 & 22.53 & 40,800 \\ 
			5 & 5 & 5 & 5 & 5 & 5 & 5 & 5 & 5 & 5 & 5 & 4 & 5 & 5 & 4 & 5 & 5 & 4 & 20.53 & 41,760 \\ 
			5 & 5 & 5 & 5 & 6 & 5 & 5 & 5 & 5 & 5 & 5 & 5 & 5 & 5 & 4 & 5 & 5 & 3 & 19.53 & 42,240 \\ 
			5 & 5 & 5 & 6 & 6 & 5 & 5 & 5 & 5 & 5 & 5 & 5 & 5 & 5 & 5 & 5 & 5 & 3 & 17.90 & 43,200 \\ 
			5 & 5 & 5 & 6 & 6 & 5 & 5 & 5 & 5 & 5 & 5 & 5 & 5 & 5 & 5 & 5 & 5 & 4 & 17.48 & 43,680 \\ 
			5 & 5 & 5 & 6 & 6 & 5 & 5 & 6 & 5 & 5 & 5 & 5 & 5 & 5 & 5 & 5 & 5 & 5 & 16.55 & 44,640 \\ 
			6 & 6 & 6 & 6 & 6 & 4 & 6 & 6 & 6 & 6 & 6 & 6 & 6 & 6 & 6 & 2 & 6 & 4 & 14.80 & 48,000 \\ 
			6 & 6 & 6 & 6 & 6 & 4 & 6 & 6 & 6 & 6 & 6 & 6 & 6 & 6 & 6 & 6 & 6 & 2 & 13.21 & 48,960 \\ 
			6 & 6 & 6 & 5 & 6 & 4 & 6 & 6 & 6 & 6 & 6 & 6 & 6 & 6 & 6 & 6 & 6 & 4 & 11.01 & 49,440 \\ 
			6 & 6 & 6 & 6 & 6 & 6 & 4 & 6 & 6 & 6 & 6 & 6 & 6 & 6 & 6 & 6 & 6 & 4 & 8.83 & 49,920 \\ 
			6 & 6 & 6 & 6 & 6 & 6 & 6 & 6 & 6 & 6 & 6 & 6 & 6 & 6 & 6 & 6 & 6 & 3 & 8.54 & 50,400 \\ 
			6 & 6 & 6 & 6 & 6 & 6 & 6 & 6 & 6 & 6 & 6 & 6 & 6 & 6 & 6 & 6 & 6 & 4 & 8.23 & 50,880 \\ 
			6 & 6 & 6 & 6 & 6 & 6 & 6 & 6 & 6 & 6 & 6 & 6 & 6 & 6 & 6 & 6 & 6 & 5 & 8.19 & 51,360 \\
			\bottomrule			
	\end{tabular}}	
\end{table*}
\renewcommand\arraystretch{1}

%


\clearpage
%


%
%
%

\end{document}